\documentclass[sn-mathphys,Numbered]{sn-jnl}

\usepackage{graphicx}%
\usepackage{multirow}%
\usepackage{amsmath,amssymb,amsfonts}%
\usepackage{amsthm}%
\usepackage{mathrsfs}%
\usepackage[title]{appendix}%
\usepackage{xcolor}%
\usepackage{textcomp}%
\usepackage{manyfoot}%
\usepackage{booktabs}%
\usepackage{algorithm}%
\usepackage{algorithmicx}%
\usepackage{algpseudocode}%
\usepackage{listings}%
\usepackage{comment}%

\def\4He{$^4$He}

\newcommand{\red}[1]{{\color{red} {#1}}}
\newcommand{\blue}[1]{{\color{blue} {#1}}}

\begin{document}

\title[Article Title]{Experimental Challenges in Determining Heat Transfer Efficiency Scaling in Highly Turbulent Cryogenic Rayleigh–B\'{e}nard Convection}
\author*[1]{P. Urban}\email{urban@isibrno.cz}
\author[1]{V. Musilov\'{a}}
\author[1]{P. Hanzelka}
\author[1]{T. Kr\'{a}l\'{i}k}
\author[1]{M. Macek}
\author*[2]{L. Skrbek}\email{ladislav.skrbek@matfyz.cuni.cz}
\affil[1]{Czech Academy of Sciences, Institute of Scientific Instruments, Kr\'{a}lovopolsk\'{a} 147, 612 00 Brno, Czech Republic}
\affil[2]{Charles University, Faculty of Mathematics and Physics, 
Ke Karlovu 3, 121 16 Prague, Czech Republic}

\date{\today}
\abstract
{Cryogenic Rayleigh–B\'{e}nard convection (RBC) at very high Rayleigh numbers (Ra) serves as a key system for understanding buoyancy-driven industrial and large scale natural flows and for testing theories of turbulent convective heat transport. 
Cryogenic helium experiments allow one to reach extremely high Ra under well-controlled laboratory conditions; however, interpretation of the resulting heat-transfer scalings remains sensitive to non-Oberbeck–Boussinesq (NOB) effects, experimental uncertainties, as well as a number of corrections that ought to be applied to raw data, including corrections for the adiabatic temperature gradient, parasitic heat leaks, or finite thermal conductivity of plates and sidewalls of RBC cells. We present an analysis of experimental uncertainties and data corrections procedures applicable to cryogenic RBC experiments, specifically to those performed in cylindrical RBC cells in Brno: 
measurement uncertainties, parasitic effects, choice of $^4$He working points in the \textit{p -- T }diagram and evaluation of relevant properties of the particular working fluid in connection with the available thermophysical property databases. In particular, our study highlights the necessity of rigorous uncertainty analysis for assessing experimental evidence suggesting either transition to the ultimate regime of RBC due to intrinsic ultimate-regime dynamics or as a manifestation of NOB effects and experimental imperfections.
}


\maketitle

\section{Introduction}
\label{sec:intro}

The Rayleigh-B\'{e}nard convection (RBC) \cite{VermaBook} occurs in a fluid layer 
confined between laterally infinite, perfectly conducting plates heated from below in a gravitational field. For an Oberbeck-Boussinesq (OB) fluid, it is 
characterized by two control parameters: the Rayleigh, $\rm{Ra}$, and the Prandtl numbers, $\rm{Pr}$. The convective heat transfer efficiency is the most robust response parameter of RBC and is described by the Nusselt number, via the $\rm{Nu = Nu(Ra; Pr)}$ dependence. 
RBC has been studied experimentally since the 19th century, typically in cylindrical cells of diameter $D$ and height $L$; the relevant additional parameter is the aspect ratio ${\rm{\Gamma}} = D/L$, which serves as a first approximation in taking into account the shape of the RBC cell. The dimensionless numbers describing RBC are:
\begin{equation}
  {\rm{Nu}} = \frac{Lq_{\rm B}}{\lambda \Delta T}\,; \quad    {\rm{Ra}}=g \frac{\alpha}{\nu\kappa}\Delta T L^3\,; \quad  {\rm{Pr}}=\frac{\nu}{\kappa}\,. 
  \label{basic}
\end{equation}
\noindent
Here $q_{\rm B}$ is the convective heat flux density, $g$ stands for the acceleration due to gravity, and $\Delta T$ is the temperature difference between the top and the bottom plates separated by the vertical distance $L$. The properties of the working fluid are characterized by the thermal conductivity, $\lambda$, and by the combination $\alpha/(\nu \kappa)$, where $\alpha$  is the isobaric thermal expansion, $\nu$  is the kinematic viscosity, and $\kappa$  denotes the thermal diffusivity.
These considerations assume an Oberbeck-Boussinesq (OB) working fluid -- a fluid with constant physical properties; only small density changes linearly dependent on temperature are accepted. In experiments, the OB conditions are never fully satisfied and this might lead to serious consequences, especially for laboratory experiments performed with gaseous working fluids in the vicinity of their equilibrium saturated vapor pressure (ESVP) curves and the critical point (CP), especially at very high $\rm{Ra}$ \cite{Elusive}. Very high $\rm{Ra}$ experiments performed at ambient temperature have typically used pressurized $\rm{SF_6}$ (Goettingen \cite{Ahlers}). Over years, several groups (Chicago \cite{Castaing1989,Wu}, Grenoble \cite{ChavannePRL,RocheNJPhys}, Eugene \cite{NatureNiemela,EugeneLSC}, Trieste \cite{Trieste1}) followed the 1975 pioneering cryogenic RBC experiment of Threlfall \cite{Threlfall} and utilized cryogenic helium gas as a working fluid with ``magic" \cite{SkrNieUr}, well known \cite{REFPROP,HEPAK,XHEPAK} and \emph{in situ} tuneable properties.

\begin{figure}
    \centering
    \includegraphics[width=0.99\linewidth]{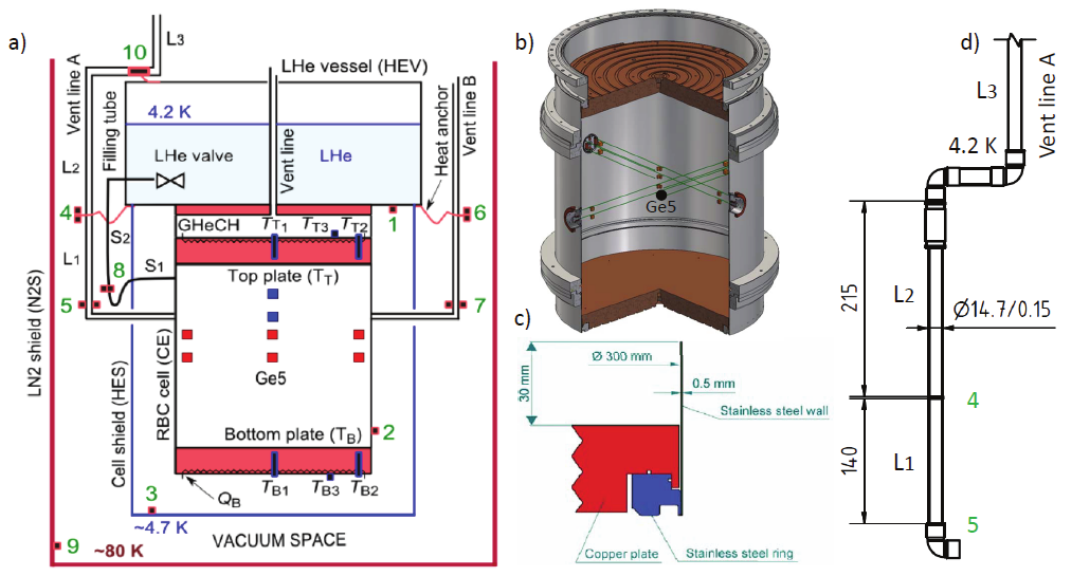}
    \caption{a: Schematic drawing of the liquid-helium (LHe) subsystem of the Brno cryostat \cite{BrnoCellRSI} with a cylindrical RBC cell of aspect ratio $\Gamma =1$. Positions 1–3 indicate locations of thermometers at the bottom of the LHe vessel, on the cell sidewall, and at the bottom of the cell shield, respectively. Positions 4, 6, and 10 indicate thermal anchoring points of the vent lines A and B to the LHe vessel. At positions 4–7, thermometers and heaters are installed for temperature monitoring and control. Position 8 marks the thermometer and heater on the 3.2 mm diameter filling tube supplying helium from the LHe vessel to the RBC cell when the cold valve inside the LHe vessel is open. The cell is mounted via the gaseous $^4$He filled heat exchange chamber (GHeCH) to the bottom of the LHe vessel. b: The RBC cell with the holders of miniature temperature sensors; the central sensor Ge5 measuring the bulk temperature and its fluctuations, $T_{\rm{Ge5}}(t)$, is highlighted. c: Construction detail of the attachment of the thin stainless-steel sidewall to the copper bottom plate (the top plate is attached analogously). A 0.5 mm gap separates the plates from the sidewall at the cell corners. Copper parts are shown in red. d: Construction detail (dimensions in mm) of the vent line A connecting the RBC cell to the cryostat flange at room temperature. 
    }
    \label{fig:conev} 
\end{figure}

The possibility of accessing a large range of Ra up to the highest Ra attainable in RBC laboratory experiments ($\approx 10^{17}$) \cite{NatureNiemela} is, however, not the only advantage of using cryogenic helium gas.
Cryogenic experiments benefit from an excellent thermal isolation thanks to the deep cryogenic vacuum surrounding the RBC cell. In addition, comparative analyses of RBC experiments show that thermal radiation within the RBC cell, although relevant at near-ambient temperatures, is negligible for cryogenic helium RBC at temperatures of a few Kelvin \cite{ThRadiation}. Favorable cryogenic properties of various construction materials such as copper or stainless steel allow for the design of the RBC cell with minimum influence of its structure on the studied convective flow. In order to utilize these advantageous considerations, our group designed and constructed an experimental cryostat \cite{BrnoCellRSI} that allows performing various types of RBC experiments, starting with steady RBC in the aspect ratio $\Gamma=1$ cylindrical cell 30 cm in diameter. Since then this apparatus has been repeatedly improved in various ways. This paper describes its current status and focuses in particular on all sources of experimental uncertainties affecting the control parameters Ra, Pr and the heat transfer efficiency Nu, Eq.~\ref{basic}.  

\begin{figure}
    \centering
    \includegraphics[width=0.99\linewidth]{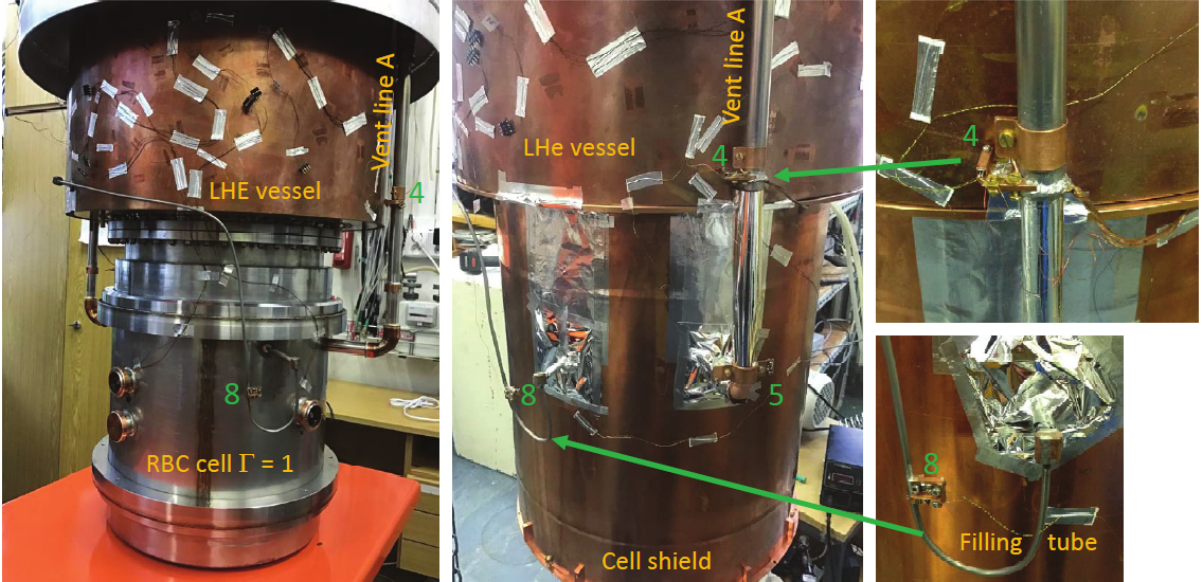}
    \caption{Photographs of the LHe subsystem of the current Brno RBC apparatus, containing the newest version of the $\Gamma=1$ RBC cell shown in Fig.~\ref{fig:conev}b. The left image shows the cell when the helium shield (HES) is removed; vent line A and the filling tube connected to the cell are visible. The cell is connected to the LHe vessel via the GHeCH. b: The same assembly with the HES installed. Positions 4, 5, and 8 indicate locations of thermometers and heaters, as defined in Fig.~\ref{fig:conev}. c: Construction detail showing the resistive heater and thermometer on vent line A at position 4. d: Detailed view of the siphon-shaped section of the filling tube with the attached resistive heater and thermometer at position 8.
    }
    \label{fig:conev1}
\end{figure}

\section{Experimental apparatus}
Schematic of the liquid-helium (LHe) subsystem of the Brno helium cryostat is shown in Fig.~\ref{fig:conev}a. The parts of the cryostat surrounding the experimental RBC cell remain the same as described in Ref. \cite{BrnoCellRSI}, except for a few important improvements described below. Over years, cylindrical RBC cells of the same diameter 30 cm and aspect ratio 1 and 2 have been developed. Both cells have been mounted via the gaseous $^4$He filled heat exchange chamber below the liquid He vessel.  Four photographs (Fig.~\ref{fig:conev1}) illustrate the LHe subsystem of the cryostat with the latest version of the $\Gamma=1$ cell without and with radiation cell shield (HES) and the current filling tube as well as the two vent lines. 
The cell shield temperature does not exceed 4.7 K, thus the radiative parasitic heat leak to the cell \cite{ThRadiation,BrnoRadCryogenics} is suppressed to less than 1\% of the lowest convective heat flux typically used in the experiment, measured within 0.2\% accuracy \cite{WarsauETC}. While near-field radiative heat transfer discussed in Ref. \cite{BrnoRadCryogenics} is not expected to occur due to the large thermal shield–cell separation, this reference provides a relevant framework for assessing and excluding such effects in cryogenic RBC experiments.

\blue{The plates are heated by resistance-wire heaters embedded in spiral grooves on their outer surfaces. The heaters (constantan wire 0.14 mm in diameter, total resistance 180 $\Omega$ at 5 K) are operated well below the maximum current of 0.2 A. The wires are wound using the bifilar method and fixed with cryogenic varnish to ensure good thermal contact \cite{BrnoCellRSI}.}

Here we describe the modifications of the current $\Gamma=1$ cell, which in recent years produced data for several published papers, notably Ref: \cite{PRL22,PRFluidsNew,SciRep}. Compared to the previous $\Gamma$ = 1 design and to the $\Gamma$ = 2 cell, the present version incorporates the following essential improvements: 
\begin{itemize}
    \item Cell flanges: The formerly demountable indium-sealed flanges were replaced by welded flanges, improving long-term mechanical stability and enabling reliable operation at pressures up to 250 kPa. 
    \item Filling tube: The filling tube was redesigned and replaced by a stainless-steel version optimized for RBC experiments. In addition, a temperature sensor with a PID (proportional–integral–derivative) controlled heater was installed on the filling tube (position 8 in Fig. 1), enabling active temperature control. 
    \item A removable holder carrying a set of miniature temperature sensors inside the cell has been developed \cite{RBCdrzakSensoru}, aiming that it influences the buoyancy-driven turbulent RBC flow of cryogenic helium gas as little as possible. The holder consists of a pair of identical feedthroughs (each with 6 glass-sealed pins), constantan wires (0.1 mm), Be bronze springs and insulating elements and enables installing up to 6 sensors simultaneously, with the option to replace the damaged ones. These holders allow obtaining a map of local temperature and its fluctuations in 12 positions inside the cell.
\end{itemize}

\section{Measured quantities and their uncertainties}
\begin{figure}
    \centering
    \includegraphics[width=0.99\linewidth]{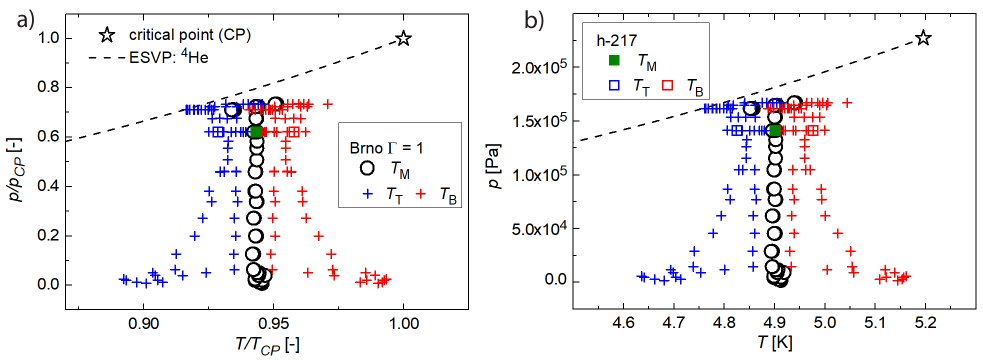}
    \caption{The $p-T$ diagram of $^4$He, shown a: in dimensionless relative units with respect to the critical pressure and temperature and b: in absolute units, with the working points (chosen far from the critical point) as used in recent $\Gamma=1$ Brno cryogenic RBC experiments. The crosses denote the corresponding temperatures of the top, $T_{\rm T}$, and bottom, $T_{\rm B}$, plates. The working point h-217 with corresponding $T_{\rm B}$ and $T_{\rm T}$ ($\rm{Ra}=4.65 \times 10^{12}$) for which the time records are shown in Fig.~\ref{fig:fluctuations} is highlighted - displayed as squares.}
    \label{fig:PhaseDia}
\end{figure}

Basic properties of confined steady-state turbulent high Ra RBC flow of OB working fluid are 
determined by dimensionless control parameters Ra, Pr and $\Gamma$, which are evaluated together with Nu, using the relationships Eq.~\ref{basic}, based on directly measured pressure and temperatures in various places of the RBC cell. These quantities are relevant to each particular working point in the pressure-temperature $p-T$ diagram of $^4$He, shown in both absolute and dimensionless form in Fig.~\ref{fig:PhaseDia}. The $p-T$ diagram includes a selected set of operating points, shown together with their corresponding 
temperatures of the top, $T_{\rm T}$, and bottom, $T_{\rm B}$, plates. We have chosen this particular choice of experimental conditions to illustrate the influence of various corrections applied to raw data on resulting Nu=Nu(Ra) scaling. We emphasize that different working points in the $p-T$ diagram in fact represent different working fluids of physical properties spanning very large intervals of values, demonstrating the flexibility of $^4$He based cryogenic fluid dynamics \cite{SkrNieUr}.

\begin{figure}
    \centering
     \includegraphics[width=0.95\linewidth]{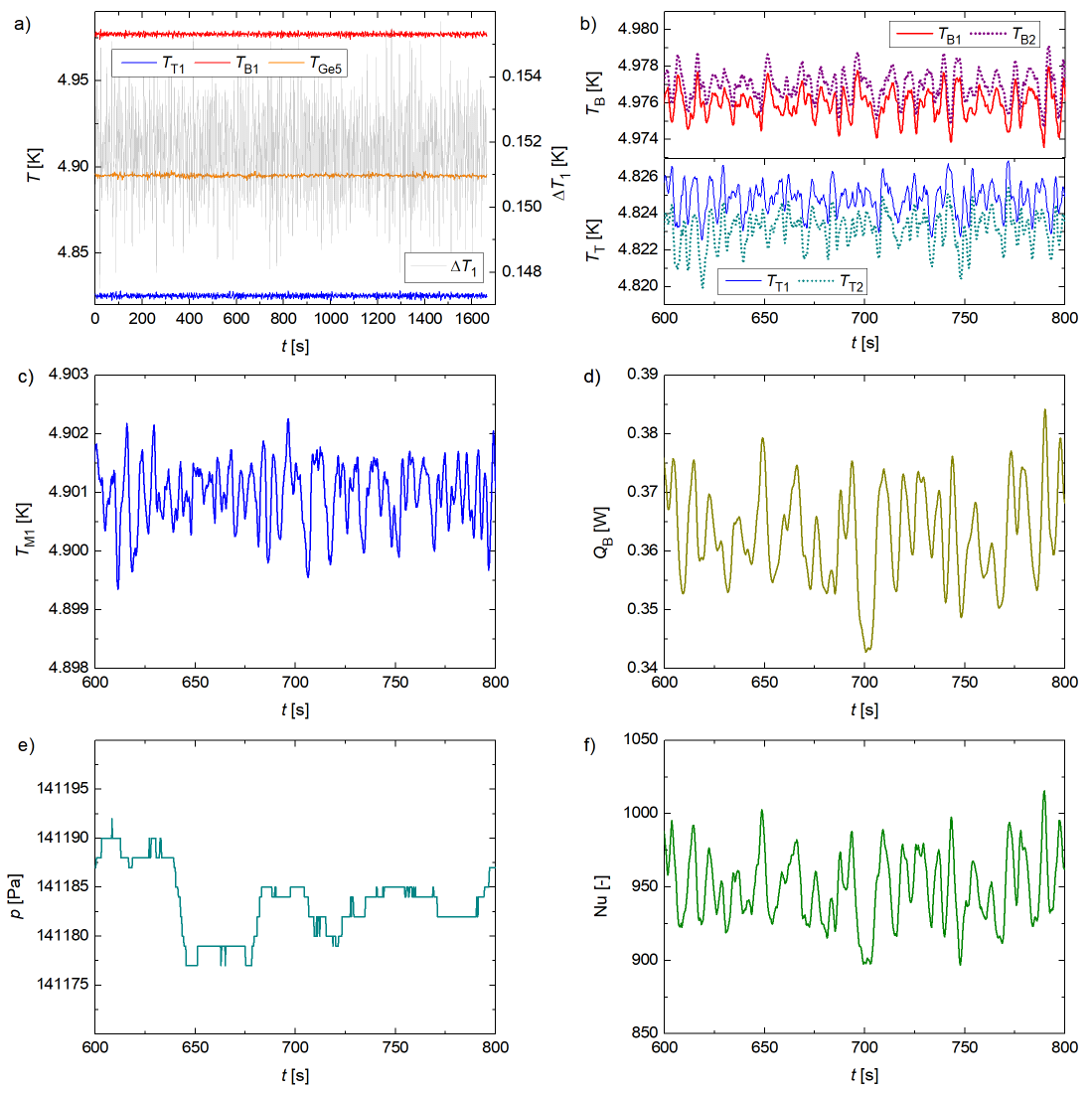}
    \caption{Examples of time series of measured and derived quantities for ${\rm{Ra}}= 4.65 \times 10^{12}$ (recorded for the working point h-217 highlighted in Fig.~\ref{fig:PhaseDia}) illustrate temporal stability and fluctuations typical for steady states of RBC. a: Time series of the temperatures recorded in the middle of the top plate,  $T_{\rm{T1}}$, bottom plate, $T_{\rm{B1}}$, central temperature $T_{\rm{Ge5}}$ (left axis), and the temperature difference $\Delta T_1=T_{\rm{B1}}-T_{\rm{T1}}$ (right axis). b: closer views on fluctuations and long-term stability of $T_{\rm{T1}}$ and $T_{\rm{B1}}$ \blue{together with temperatures $T_{\rm{T2}}$ and $T_{\rm{B2}}$ measured at the edge of plates, with synchronous changes of middle and edge temperatures visible.} c: The mean plate temperature $T_{\rm{M1}}=(T_{\rm{B1}}+T_{\rm{T1}})/2$. d: The record of heating power $Q_{\rm B}$ into the bottom plate. e: The record of the cell pressure $p$. f: The instantaneous Nusselt number, Nu. The displayed time records represent one of operating points used to construct the Nu(Ra) dependence. For this particular Ra, the time-averaged density and pressure of the working $^4$He gas were $\rho=19.9\,$kg~m$^{-3}$, $p = 141.189$ kPa.
    }
    \label{fig:fluctuations}
\end{figure}

Of primary importance is the dimensionless heat transfer efficiency, i.e., the Nu(Ra) dependence. The thermophysical properties of $^4$He required in Eqs. (1) are obtained from one of available helium property databases: REFPROP \cite{REFPROP}, HEPAK \cite{HEPAK} and XHEPAK \cite{XHEPAK}), at the working state point defined by the pressure $p$ and the mean fluid temperature $T_{\rm M} = (T_{\rm B} + T_{\rm T}) / 2$. 
The temperature difference $\Delta T = T_{\rm B} – T_{\rm T}$ enters both Ra and Nu. The convective heat flux needed for the evaluation of Nu is determined from the electrical heating power $Q_{\rm B}$ supplied to the bottom plate, with possible corrections for parasitic losses. It is crucial to know the accuracy of these quantities, extracted from the experiment. The analysis of the Nu(Ra) dependence in this study is based on the HEPAK \cite{HEPAK} database, with the exception of Fig.~\ref{fig:All}a.

\blue{Examples of time series of measured and derived quantities for ${\rm{Ra}} = 4.65 \times 10^{12}$, illustrating the typical temporal stability and fluctuations in steady states of turbulent RBC, are presented in Fig. 4.}

\subsection{Pressure and heating power measurement} 

\textbf{The gas pressure}, $p$, inside the cell is measured using the Baratron 690A 53T RB transducer with an accuracy of about 0.08\%.  Pressure sensor is positioned above the cell at the room temperature part of the vent line, measuring thus a tiny bit smaller value than the pressure, $p$, at the center of the RBC cell. For the measurements  presented here the pressure correction on hydrostatic pressure would be, however, less than the uncertainty of 0.08\% in pressure measurement and can be neglected. Similarly, the hydrostatic pressure difference between the top and bottom of the cell is ~0.07\% or less and can be neglected, too.

\textbf{Heating power}, $Q_{\rm B}$, supplied to the bottom plate is obtained from four-wire measurements
of voltage and current using two Keysight 34460A multimeters, with an
overall accuracy better than 0.2\%. 

Typical fluctuations and long-term stability of $p(t)$
and $Q_{\rm B}(t)$ for steady RBC under the working conditions (h-217) highlighted in Fig. 3
are illustrated in Fig. 4.



\subsection{Temperature measurement:} 

The temperatures of the top and bottom copper plates, $T_{\rm T}$ and $T_{\rm B}$, are measured by four calibrated germanium sensors Lake Shore GR-200A-1500-1.4B. According to the Calibration Certificates issued by PTB Berlin, the standard uncertainty is 1 mK of the calibration in the range from 1.78 K to 10.45 K for three of them \blue{($T_{\rm{T2}}$, $T_{\rm{B1}}$, $T_{\rm{B2}}$)} and 3 mK for the last one $(T_{\rm{T1}})$. 
These four calibrated sensors are embedded in the middle $(T_{\rm{T1}},\,T_{\rm{B1}})$ and at the edge $(T_{\rm{T2}},\,T_{\rm{B2}})$ of both Cu plates. In the experiment, the sensor resistances are read by an LS340 temperature controller. Its resistance inputs were calibrated over the range 0–3000~$\Omega$ using a precision MEATEST resistance standard. Calibration curves were established and applied to correct the LS340 readings, thereby improving the accuracy of the measured resistance values beyond the nominal accuracy specified by the LS340 manufacturer (Lake Shore Cryotronics). The MEATEST standard provides an accuracy approximately one order of magnitude higher than that of the LS340.

\blue{The resistances measured by the LS340 were converted to temperature using the calibration curves of the germanium sensors, which have a sensitivity of $|dR/dT | \approx 862\, \Omega$/K at 5 K. The resulting uncertainty from the controller’s resistance measurement ($\approx 0.2 \,\Omega$, $\approx 0.23\,$mK) is rather small compared to the sensor’s calibration uncertainty of $\pm 1\,$mK.}

For analyzing steady-state RBC, the time-averaged readings $(T_{\rm{T1}},\,T_{\rm{B1}})$, $(T_{\rm{T2}},\,T_{\rm{B2}})$ are used to construct the following temperatures: mean central temperature, $T_{\rm{M1}}= (T_{\rm{B1}} + T_{\rm{T1}})/2$ and temperature difference, $\Delta T_1= T_{\rm{B1}} - T_{\rm{T1}}$, mean edge temperature $T_{\rm{M2}}= (T_{\rm{B2}} + T_{\rm{T2}})/2$ and temperature difference $\Delta T_2= T_{\rm{B2}} - T_{\rm{T2}}$, averaged top $T_T=(T_{\rm{T1}} + T_{\rm{T2}})/2$ and bottom $T_{\rm B}=(T_{\rm{B1}} + T_{\rm{B2}})/2$ temperatures and the mean temperature $T_{\rm M}=(T_{\rm T} + T_{\rm B})/2=(T_{\rm{M1}} + T_{\rm{M2}})/2$. 

\begin{figure}[h]
    \centering
     \includegraphics[width=0.99\linewidth]{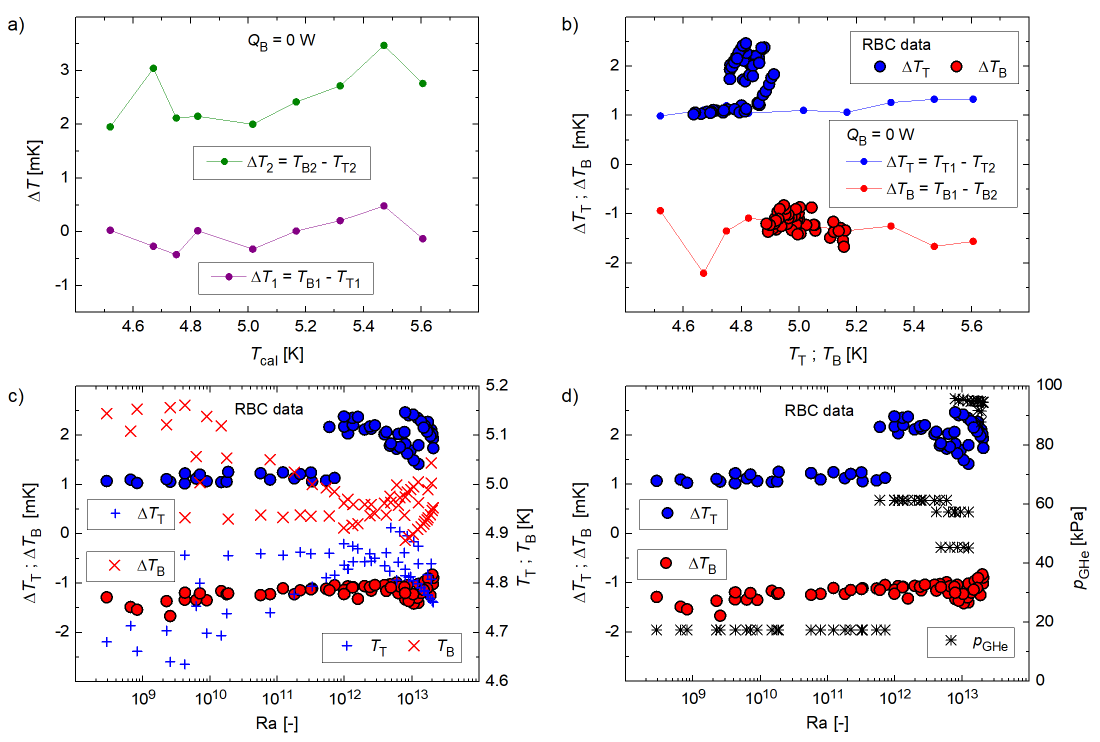}
    \caption{a: Measured temperature difference between the top and bottom plates during the calibration of the internal TTR-G miniature germanium sensors ($Q_{\rm B}$ = 0 W), recorded by four calibrated GR-200A-1500-1.4B reference sensors using the LS340 temperature controller. b: Temperature differences between the central and edge sensors on the top and bottom plates as a function of plate temperatures recorded during the calibration of the internal TTR-G miniature germanium sensors ($Q_{\rm B}$ = 0 W) and during the RBC experiment (RBC data). The sensors are embedded in the middle $\Delta T_1=T_{\rm{B1}}-T_{\rm{T1}}$ (blue) and at the edge of plates $\Delta T_2=T_{\rm{B2}}-T_{\rm{T2}}$ (red). c: Temperature differences between the central and edge sensors on the top and bottom plates as a function of all measured Ra. The right vertical axis shows the absolute temperatures of the top and bottom plates, $T_{\rm T}$ and $T_{\rm B}$, defined as averages of the two sensors on each plate. d:  The same temperature differences, with the right vertical axis showing the pressure $p_{\rm{GHe}}$ in the helium heat-exchange chamber positioned between the RBC cell below and the liquid helium vessel above.}
    \label{fig:plateDeltaT}
\end{figure}

\vspace{3mm}
\textbf{Calibration procedure for internal miniature TTR-G germanium sensors}. These fast response sensors, located in the cell interior, have been calibrated against the GR-200A-1500-1.4B reference sensors embedded in the cell plates.
A set of nine calibration temperature points was chosen in the range 4.5–5.6~K (see the horizontal axes in the top panels of Fig.~\ref{fig:plateDeltaT}). The RBC cell was filled with low-density $^4$He gas ($\rho = 3.6$~kgm$^{-3}$), and the gaseous $^4$He heat exchange chamber  (GHeCH) was pumped down to 4.5~kPa to minimize temperature fluctuations of $T_{\rm T}$ caused
by helium convection in the GHeCH. \blue{The cell was then slowly heated by applying power $Q_{\rm B}$ to the bottom-plate heater until convection of the $^4$He gas raised the top-plate temperature to approximately 50~mK above the selected calibration point $T_{\rm cal}$. This procedure ensured thorough turbulent mixing of the cell bulk 
(for this working point, the onset of convection practically equals adiabatic gradient, Sec. 4.2, which is lower than 1 mK in presented experiments, see Fig. 7), so that initially $T_{\rm B} > T_{\rm M} > T_{\rm T} > T_{\rm cal}$. The bottom heating power $Q_{\rm B}$ was then switched off, and the top-plate temperature $T_{\rm T}$ was stabilized at $T_{\rm cal}$ using PID control. The system was subsequently allowed to relax to the state without convection (Figs. 5a, b) followed by very slow equilibration by thermal molecular diffusion. The reference plate temperatures and the resistance values of the TTR-G sensors were recorded. For each calibration point, the calibration temperature $T_{\rm cal}$ was determined as the average of the top- and bottom-plate temperatures, $T_{\rm T}$ and $T_{\rm B}$.}


\vspace{3mm}
\textbf{Temperature homogeneity of the plates.} The top and bottom plates of our RBC cells have been made of 28 mm thick oxygen-free (OFHC) copper (RRR = 290). The thermal conductivity of the plates is very high, about 2000 Wm$^{-1}$K$^{-1}$ at 5~K \cite{EPLus}. The plates have a smooth, finely turned inner surface. The design of the heaters glued into the spiral grooves milled on the external sides of the plates \cite{BrnoCellRSI} ensures better than 1~mK temperature homogeneity on the internal plate surfaces, assuming uniform supply or removal of heat. The plate temperatures $T_{\rm T}$ and $T_{\rm B}$ are stabilized by PID control using the $T_{\rm{T1}}$ and $T_{\rm{B1}}$ sensors on the top and bottom plates, respectively, via the Lake Shore 340 controller. 
\blue{Despite that, Fig.~\ref{fig:plateDeltaT} (b-d) shows that convection in the RBC cell together with convection in the gaseous He exchange chamber above it degrade the plate temperature homogeneity, resulting in deviations larger than 1~mK.
The contribution of convection in the exchange chamber increases with increasing
pGHe, see Fig. 5d.}

\blue{\textbf{Uncertainty in Nu(Ra) scaling due to uncertainty in pressure and mean temperature, $T_{\rm M}$.} The uncertainties in pressure and temperature values generate uncertainty in $^4$He properties read from databases of thermophysical properties (HEPAK [15], XHEPAK [16] or REFPROP[14]). In measurements presented
in Figs. 3, 6 and 9, corresponding uncertainty in the Nu/Ra$^{1/3}$ value (Fig. 6)
decreases with pressure and is less than 0.8\%. Similarly, the uncertainty in Ra is of 2\%
or less. The uncertainty in Nu/Ra$^{1/3}$ due to the uncertainty in mean temperature $T_{\rm M}$
is less than 0.5\% per mK. Corresponding uncertainty in Ra is about 1\% per mK and
less. Uncertainties in Nu/Ra$^{1/3}$ and Ra are partly correlated, having opposite signs in the presented experiments.}


The total uncertainty in Nu/Ra$^{1/3}$ arises from uncertainties in the mean temperature $T_M$ ($\approx 2$ mK), pressure $p$ (0.08 \%), temperature difference $\Delta T$ ($\approx 2$ mK), and heat input $Q_{\rm B}$ ($\approx 0.2$ \%). The dominant contribution comes from the uncertainty in $\Delta T$, while uncertainties in $T_{\rm M}$, $p$, and $Q_{\rm B}$ have much lesser effect on the Nu/Ra$^{1/3}(\rm{Ra})$ scaling.

\section{Discussion}

It is instructive to discuss the influence of various factors and corrections on a typical real data set measured in the described above $\Gamma=1$ cylindrical RBC cell, spanning several orders of high Ra. In this study, we do not discuss the underlying physics, which will be published elsewhere; we merely show the influence of the discussed uncertainties and corrections to the measured data.

\subsection{Corrections to raw data}

\begin{figure}
    \centering
     \includegraphics[width=0.99\linewidth]{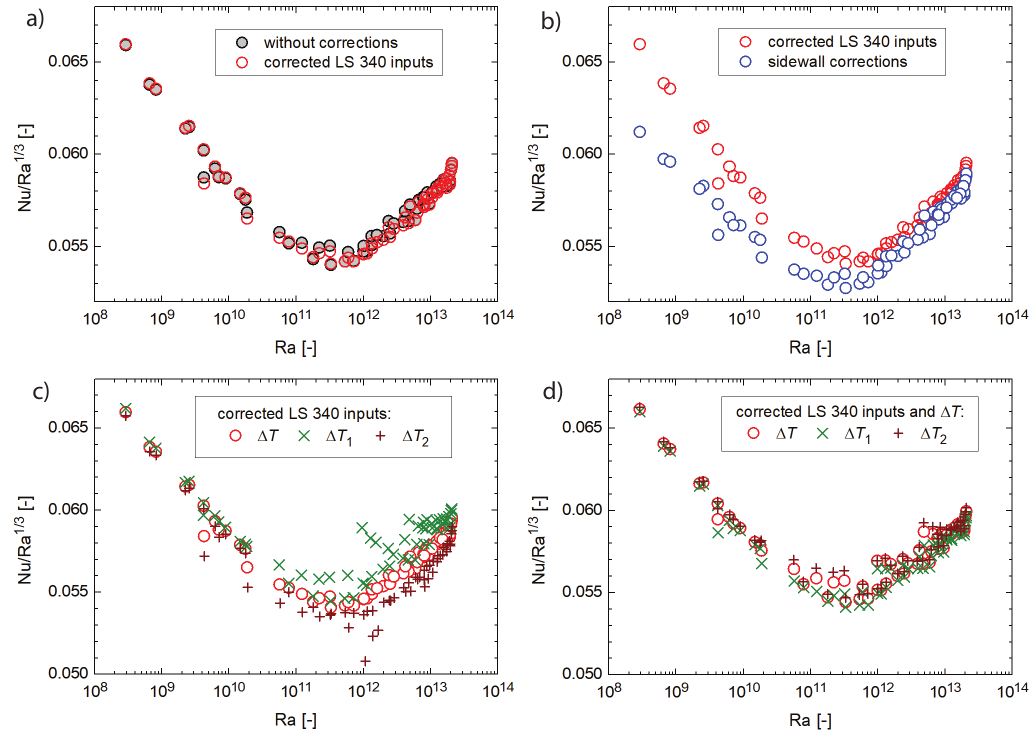}
    \caption{Influence of corrections discussed in the text on the evaluated Nu/Ra$^{1/3}$ dependence. a: Influence of re-calibration of the LS340 temperature controller against the MEATEST resistance standard on  Nu/Ra$^{1/3}(\rm{Ra})$ versus Ra dependence, evaluated for the measurements in working points shown in Fig.~\ref{fig:PhaseDia} based on $T_{\rm M}=(T_{\rm B}+T_{\rm T})/2$, $\Delta T= T_{\rm B}-T_{\rm T}$, and HEPAK \cite{HEPAK} database. Hereafter, the corrected against the MEATEST resistance standard data are considered as raw data. b: Influence of sidewall corrections, applied to all raw data points. c: Influence of plate temperature inhomogeneity on the evaluated Nu/Ra$^{1/3}(\rm{Ra})$ dependence. \blue{d: the same, but with $\Delta T$, $\Delta T_1$ and $\Delta T_2$ corrected to the offsets determined in the calibration at $p=0.14\,$MPa and $T_{\rm M}=4.9\,$K at $Q_{\rm B} = 0\,$W.}
          }
    \label{fig:Corr}
\end{figure}

Despite the cryogenic cell is constructed bearing in mind its minimal influence on the turbulent RBC flow under study, even under assumption of ideal OB working fluid (see later for NOB corrections) various corrections ought to be considered and some of them applied to raw data such as just discussed. We illustrate their influence on evaluated Nu(Ra) scaling (we use the more sensitive, compensated Nu/Ra$^{1/3}(\rm{Ra})$ versus Ra plot) in Fig.~\ref{fig:Corr}. 

\vspace{3mm}
\blue{\textbf{Corrected LS340 inputs.} Calibration of the resistance values measured by
the LS340 temperature controller against the MEATEST resistance standard was
performed. The required corrections (not caused by the RBC flow under study)
are straightforward and affect the evaluated Nu/Ra$^{1/3}$ versus Ra dependence only
marginally, as documented in Fig. 6a. Hereafter this correction is always applied and
this data set is understood as \textbf{raw data}.}

\vspace{3mm}
\textbf{Finite heat conductivity and heat capacity of plates.}
Thanks to the high heat conductivity and overall quality of our Cu plates, corrections related to the physical properties of the plates as suggested in Ref.~\cite{ChillaPlates} to our raw data are not applied.

\vspace{3mm}
\textbf{Sidewall corrections.}
The parasitic heat fluxes caused by the finite thermal conductivity of the sidewalls in our RBC cells are minimized by using very thin (0.5~mm) stainless steel of relatively low thermal conductivity. This thin sidewall is used for the parts of the RBC cell adjacent to plates, i.e., where it matters (Fig. 1c). Sidewall corrections play a significant role at low Ra $ \lessapprox 10^{11}$. In Ref. \cite{BrnoPRL1} we demonstrated that the standard wall correction applied to raw RBC data result in a consistent collapse of Nu/Ra$^{1/3}$ versus Ra dependencies at low Ra across three independent cryogenic experiments conducted in cylindrical cells in Grenoble, Trieste and Brno, as well as \cite{SkrNieUr} with cryogenic $^3$He RBC experiments \cite{Meyer01}. Influence of sidewall corrections decreases with increasing Ra and have little impact at high Ra $ > 10^{11}$, where the scaling law of Nu/Ra$^{1/3}$ versus Ra is not altered appreciably. Influence of sidewall corrections, applied to all data points of the raw data, is illustrated in Fig.~\ref{fig:Corr}b. 

\vspace{3mm}
\blue{\textbf{$\Delta T$ correction (offset and adiabatic gradient contribution).} A significant correction concerns the temperature difference $\Delta T$ between the plates, which is the dominant source of uncertainty in the evaluated Nu and Ra. We aimed to decrease the uncertainty in the smallest values of $\Delta T = T_{\rm B}-T_{\rm T}$ originating from uncertaintees in $T_{\rm B}$ and $T_{\rm T}$.
Similarly to TTR-G sensors calibration procedure (Sec.~3.2), we read differences (offsets) $\Delta T_1= -0.8\,$mK and  $\Delta T_2= + 2.8\,$mK at $Q_{\rm B}=0$ for fluid state $p=0.14\,$MPa and $T=4.9\,$K and a pressure of $\approx 60\,$kPa of gaseous He in the heat exchange chamber where the data set h-217 (${\rm{Ra}}=4.65 \times 10^{12}$) was obtained; Figs. 3 and 4. These values of offsets are more appropriate with respect to the highest Ra measurements.
The reading was done after ceasing of convection within the cell (Sec.~3.2). Owing to the very long relaxation times of following equilibration by means of molecular diffusion only, the read values of $\Delta T_1$ and $\Delta T_2$ at $Q_{\rm B} = 0$ still include the remnant adiabatic temperature gradient $\Delta T_{\rm{ad}} \approx 0.85\,$mK (Sec. 4.2, Fig. 7) calculated for fluid used for this calibration point.

Fig. 6c compares Nu/Ra$^{1/3}$ vs Ra obtained using the different $\Delta T$ definitions
(Sec. 3.2), showing systematic deviations arising from the $\Delta T$ offsets. In Fig. 6d, subtraction
of the corresponding offsets determined from the TTR-G calibration at $Q_{\rm B} = 0$ ($\Delta T_1 =-0.8\,$mK and $\Delta T_2 = 2.8\,$mK) leads to a consistent collapse of all data sets ($\Delta T_1$, $\Delta T_2$ and $\Delta T$) onto a single curve.
The dependence $\Delta T_{\rm{ad}}(\rm{Ra})$ (see Fig. 7) is not included here. The full correction $\Delta T_{\rm{ad}}(\rm{Ra})$ is embodied in Nu(Ra) dependencies in Fig. 9.}

\begin{figure}
    \centering
     \includegraphics[width=0.99\linewidth]{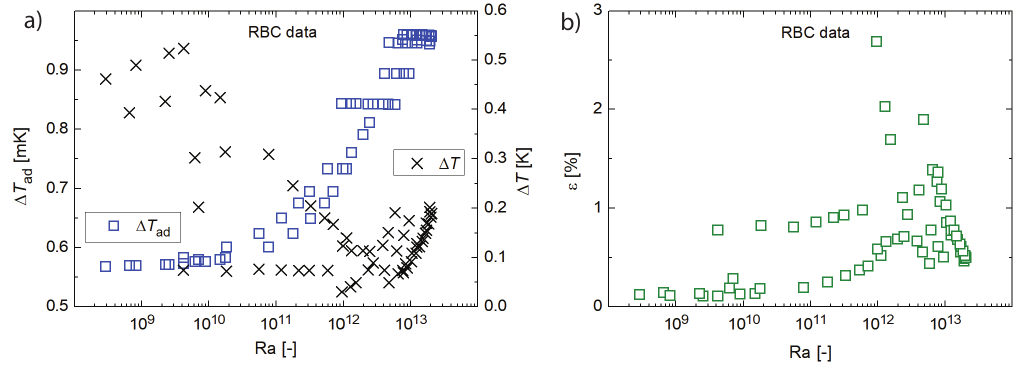}
    \caption{a: Temperature difference due to adiabatic temperature gradient, $\Delta T_{\rm{ad}}$, plotted together with the temperature difference between plates $\Delta T$. b: The relative adiabatic temperature gradient $\varepsilon=\Delta T_{\rm{ad}}/\Delta T$ plotted versus Ra. The displayed data serve as a typical example of $\Delta T_{\rm{ad}}$, calculated for experimental conditions specified in the $p-T$ diagram, Fig.~\ref{fig:PhaseDia}. }
    \label{fig:AdVera}
\end{figure}

\subsection{Adiabatic temperature gradient}

Hydrodynamic similarity of RBC, based on dimensionless numbers Ra and Pr, is inherently connected with Oberbeck-Boussinesq approximation in equations describing free convection. Among others, this approximation neglects effect of hydrostatic pressure on convection. It is possible to evaluate this effect in RBC by introduction the temperature difference, $\Delta T_{\rm{ad}}$, due to the adiabatic thermal gradient \cite{Tritton} as

\begin{equation}
    \frac{\Delta T_{\rm{ad}}}{L}=\frac{g \alpha T}{c_{\rm p}}\,,
\end{equation}
where $c_{\rm p}$ is the specific heat at constant pressure, and to correct experimental values of Nu and Ra. 
\begin{equation}
    {\rm{Ra_{corr}}}=\frac{g \alpha (\Delta T- \Delta T_{\rm{ad}})L^3}{\nu \kappa}={\rm{Ra}}(1-\varepsilon);\quad \varepsilon=\frac{\Delta T_{\rm{ad}}}{\Delta T}\,;
    \label{eq:Vera1}
\end{equation}
\begin{equation}
     {\rm{Nu_{corr}}} =\frac{L}{\lambda} \frac{q_{\rm B}-\lambda \Delta T_{\rm{ad}}/L}{\Delta T-\Delta T_{\rm{ad}}}
={\rm{Nu}} \left( \frac{1}{1-\varepsilon} \right) -\frac{\varepsilon}{1-\varepsilon}\,,
\label{eq:Vera2}
\end{equation}
where $q_{\rm B}= Q_{\rm B}/S$ denotes the heat flux applied per unit area of the bottom plate of area $S$.  For $\Delta T_{\rm{ad}}/\Delta T \ll {\rm{Nu}}$, which is the case of our experiment, see Fig.~\ref{fig:AdVera},
\begin{equation}
   {\rm{Nu_{corr}}}\approx {\rm{Nu}} \left(\frac{1}{1-\varepsilon}\right) \rightarrow \frac{{\rm{Nu_{corr}}}}{{\rm{Ra_{corr}}}^\gamma}= \frac{{\rm{Nu}}}{{\rm{Ra}}^\gamma} \left (\frac{1}{1-\varepsilon}\right )^{1+\gamma} \,,
   \label{eq:Vera3}
\end{equation}
where $\gamma$ stands for the scaling exponent, Nu=Nu(Ra)$^\gamma$.
If also other Boussinesq approximations are fulfilled, these corrections are exact. Relations \ref{eq:Vera1}, \ref{eq:Vera2} and \ref{eq:Vera3} have been approved in developed turbulence (Ra up to $10^7-10^8$) in fluids not far from critical points, specifically in cryogenic experiment with gaseous $^3$He \cite{Meyer01} and in 3D simulations of van der Waals fluid \cite{Accary}, to give some examples. \blue{The correction of $\Delta T$
for the adiabatic temperature gradient, commonly applied to raw data in cryogenic
RBC experiments, is included in the plots of Nu(Ra) dependencies in Fig. 9. }

\subsection{Parasitic heat flux analysis.}
Parasitic heat fluxes to and from the RBC cell directly affect the accuracy of the RBC experiment. For the relevant operating regime they must therefore be carefully quantified. The total parasitic heat flux into the convection cell consists of the following contributions:

•	heat flux $Q_{\rm{L1}}$ through the segments L1 of the vent lines A and B (Fig. 1a),

•	heat flux $Q_{\rm{S2}}$ through the filling tube connecting the valve inside the LHe vessel with the RBC cell,

•	heat flux through the heater current leads, including Joule heating generated in the leads ($Q_{\rm{H1}}$),

•   heat flux through the phosphor-bronze lead wires of the miniature temperature sensors inside the cell ($Q_{\rm{Ge}}$),

•	radiative heat flux $Q_{\rm{R1}}$ between the LHe shield and the cell, and

•	heat exchange via residual gas.

\vspace{3mm}
\textbf{Convection cell vent lines A and B.}
The RBC cell is equipped with two identical vent lines, A and B, each divided into three segments (L1–L3), connected by copper thermal anchors to the liquid helium (LHe) vessel at 4.2 K (Fig.~\ref{fig:conev}d). The vent lines are made of stainless-steel tubing (inner diameter 14.4 mm, wall thickness 0.15 mm) filled with static helium gas at the same pressure as in the RBC cell. The vent line segments are interconnected by high-thermal-conductivity Cu couplings. The lower end of segment L1 terminates in a copper elbow soldered to the thick-walled stainless-steel neck of the cell sidewall. 
The ventilation lines are exposed to radiative heat flux from the shield at liquid nitrogen temperature $T_9 \approx 80$ K. To minimize parasitic heat flow into the cell, the upper end of L1 segment is thermally anchored to the bottom of the LHe vessel at 4.2 K. At pressures above $\approx 100$~kPa this anchoring (and also at position 10, Fig. 1a) may lead to helium condensation, therefore, the thermal anchors and adjacent couplings are equipped with temperature sensors and heaters, allowing temperature control above the condensation threshold if required. In addition, the temperature along segment L1 is controlled to ensure that it increases away from the cell, thereby suppressing natural convection (so-called “chimney effect”) in the vent lines \cite{WarsauETC}.

\begin{figure}
    \centering
     \includegraphics[width=0.99\linewidth]{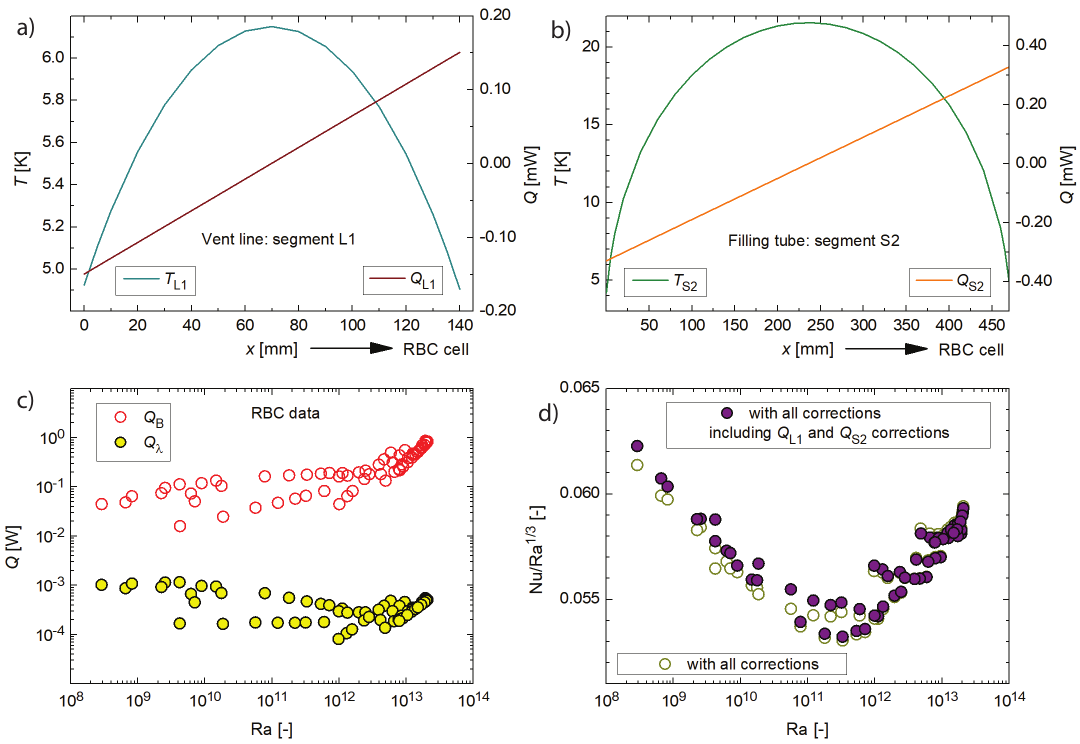}
    \caption{Calculated temperature profiles and heat flows; a:  along segment L1 of the RBC cell vent line 140 mm in length, showing temperature $T_{\rm{L1}}$ and heat flow $Q_{\rm{L1}}$; b: along segment S2 of the filling tube 470 mm in length, showing temperature $T_{\rm{S2}}$ and heat flow $Q_{\rm{S2}}$. The horizontal axes represent the position along each segment. Calculations were performed using the KRYOM 3.3 software \cite{R1} for prescribed boundary temperatures and radiative heating conditions. c: Heating power $Q_{\rm B}$ applied to the bottom plate for individual working points of the measured Nu(Ra) dependence, together with the diffusive heat flow $Q_\lambda$, their ratio defines the Nusselt number, Nu. d: Compensated plots Nu/Ra$^{1/3}(\rm{Ra})$ versus Ra, showing fully corrected data (see Fig.~\ref{fig:All}) with additional corrections for parasitic heat fluxes $Q_{\rm{L1}}$ (vent lines A and B) and $Q_{\rm{S2}}$ (filling tube).}
    \label{fig:Hanzelka}
\end{figure}

To evaluate the parasitic heat flux $Q_{\rm{L1}}$, we analyze the thermal balance of segment L1. The calculation is shown for vent line A, with identical results applying to vent line B. The temperature profile and the corresponding heat flux into the cell were calculated using the KRYOM 3.3 software \cite{R1,R2} for prescribed boundary temperatures $T_4 = 4.925$~K and $T_5 = 4.905$~K and for radiative heating from the shield at $T_9 = 80$~K. The results are shown in Fig.~\ref{fig:Hanzelka}a, yielding $Q_{\rm{L1}} = 0.15$~mW.

\vspace{3mm}
\textbf{Parasitic heat load via the filling tube.}
The RBC cell is filled through a liquid-helium (LHe) filling tube connecting the RBC cell to a valve located at the bottom, inside the LHe vessel. The filling tube consists of two segments, S1 and S2. Segment S1 is a stainless-steel tube (outer diameter 6 mm, wall thickness 0.3 mm, total length 70 mm), partially located inside the helium thermal shield (HES) at $T_3 = 4.7$~K. The part protruding from the HES is wrapped with multilayer Mylar insulation to reduce radiative heat load from the thermal alluminium shield anchored to liquid nitrogen vessel at $T_9 = 80$~K and is connected to the segment S2 by a Cu coupling.
Segment S2 is a stainless-steel capillary (outer diameter 3.2 mm, wall thickness 0.2 mm, total length 470 mm) shaped as a loop acting as a siphon to suppress natural convection. A copper block soldered to the outer surface of the capillary, located 130 mm from the S1–S2 coupling, is equipped with a temperature sensor ($T_8$) and a PID-controlled heater for active temperature control.

\begin{figure}
    \centering
    \includegraphics[width=0.99\linewidth]{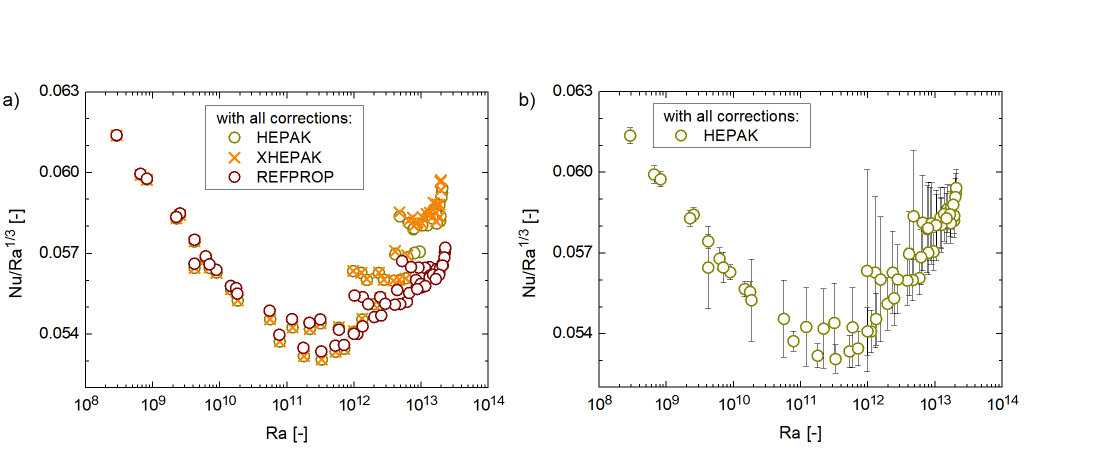}
    \caption{Compensated plots of Nu/Ra$^{1/3}(\rm{Ra})$ versus Ra. a: 
    Data with all corrections applied using thermophysical properties of $^4$He from the HEPAK, XHEPAK, and REFPROP databases, illustrating the influence of the thermophysical property databases on the evaluated Nu/Ra$^{1/3}$(Ra) dependence. b: Fully corrected data (HEPAK) with error bars based on the uncertainties of the measured quantities (Section~3).}
    \label{fig:All}
\end{figure}

The parasitic heat load into the RBC cell via the filling tube was evaluated by analyzing the thermal balance of segment S2. The temperature profile along the capillary and the corresponding heat flux into the cell were calculated using the KRYOM
3.3 software. The resulting temperature profile and heat flux are shown in Fig.~\ref{fig:Hanzelka}b. The resulting value $Q_{\rm{S2}} = 0.33$~mW corresponds to the calculation for the boundary temperatures of 4.2 K at the thermal anchor, 4.905 K at the cell, and radiative heating from the alluminium liquid nitrogen shield at $T9 = 80$~ K.

\vspace{3mm}
\textbf{Heat flux through heater current leads.}
The resistance heater in the bottom copper plate of the convection cell is powered by two copper current leads (diameter 0.09 mm). The leads are thermally anchored to the cell shield at $T_3 = 4.7$~K and to the bottom plate (total length 200 mm), whose temperature $T_{\rm B}$ varies between 4.9~K and 5.2~K. The heat flux through the current leads, including Joule heating  (the maximum current of 0.07 A was used for the data presented here), was calculated using the VVV-3 code \cite{R3}, which accounts for conduction, Joule dissipation, and radiation in vacuum. The resulting heat load $Q_{\rm{H1}}$ is below 0.03 mW in the whole temperature range and is therefore negligible.

\vspace{3mm}
\textbf{Heat flux through phosphor-bronze sensor leads ($Q_{\rm{Ge}}$).}
The heat flux through the phosphor-bronze lead wires of the miniature temperature sensors was evaluated for the section between the bottom of the LHe vessel (4.2 K) and the convection cell (4.905 K). The sensors are connected via Lake Shore WQT-36 twisted phosphor-bronze wires (diameter 0.127 mm, the average wire length is approximately 400 mm), thermally anchored at the LHe vessel bottom and led beneath the cell shield (4.7 K) to the cell feedthroughs, where they are attached. All 12 
sensors are operated in a four-wire configuration with an excitation current of 1~$\mu$A. The heat flux through the leads was calculated using the VVV-3 software code \cite{R3}, yielding a total heat load  $-8 \times 10^{-3}$~mW, which is negligible.

\vspace{3mm} 
\textbf{Radiative heat load between the cell and the HES.}
The HES is made of copper (RRR = 40, wall thickness 1 mm) and is thermally anchored to the bottom of the LHe vessel at 4.2 K, the maximum measured shield temperature is $T_3 = 4.7$~K. The radiative heat flux $Q_{\rm{R1}}$ between the HES and the convection cell at $T=4.905$~K was calculated assuming an effective emissivity of 0.5 (pessimistic estimate assuming the mostly rough stainless-steel cell sidewall). The resulting radiative heat load is $-0.18 \times 10^{-4}$~mW and is therefore negligible.

\vspace{3mm}
\textbf{Heat exchange via residual gas.}
At the operating pressure around the convection cell (below $10^{-6}$~Pa), heat transport by residual gas is negligible compared to other heat-load contributions. This assumption is supported by the agreement between measured thermal characteristics of the nearly identical NMR cryostat and calculations based on established cryostat thermal models \cite{R2}. Residual-gas heat transport can therefore be neglected.

\vspace{3mm}
\textbf{Influence of the choice of $^4$He property database on Nu(Ra) dependence.} We compare the  
Nu/Ra$^{1/3}(\rm{Ra})$ versus Ra dependencies evaluated using three databases HEPAK \cite{HEPAK}, XHEPAK \cite{XHEPAK}, and REFPROP \cite{REFPROP}, see Fig.~\ref{fig:All}a. It is well known that the helium properties evaluated using HEPAK and XHEPAK databases are almost identical, except in the close vicinity of the critical point; this, however, does not apply to the data series presented in this work. Regarding the ESVP curve, the temperature deviation between HEPAK and REFPROP is -1.7~mK at 100 kPa and 4.1~mK at 220~kPa near the critical point. The deviation changes sign around 4.9~K, with the smallest differences occurring in the temperature range relevant to the data presented here.

\vspace{3mm}
\textbf{Fig.~\ref{fig:All}b summarizes our analysis}, showing the influence of corrections to the compensated plots of Nu/Ra$^{1/3}(\rm{Ra})$ versus Ra. Parasitic heat fluxes to the cell via the vent lines A and B ($Q_{\rm{L1}}$) and the filling tube ($Q_{\rm{S2}}$), shown in Fig. 8, are not included,  as the temperature profiles along these elements vary for each measured point of the Nu(Ra) dependence. The calculations of $Q_{\rm{L1}}$ and $Q_{\rm{S2}}$ were performed under pessimistic boundary conditions to illustrate their maximum possible impact. In addition, the heat flux $Q_{\rm{S2}}$ is reduced by the thin-walled stainless-steel tube of segment S1, which provides a large thermal resistance. Consequently, the evaluated values of $Q_{\rm{L1}}$ and $Q_{\rm{S2}}$ represent rather conservative, pessimistic upper estimates. \blue{We emphasize that the displayed error bars are absolute, originating mainly from the uncertainty, up to 2 mK (see Sec. 3.2), in $\Delta T$. Thanks to $\Delta T$ corrections described in Sec.~4.1 we argue that at high Ra for any particular data series measured at constant pressure (such as that highlighted by squares in Fig.~3b) the relative errors among individual data points are much smaller.}

\section{Conclusions}
We have discussed various aspects of cryogenic experiments aimed to investigate RBC flows at very high Ra using the experimental cryostat in Brno. They include the construction details of the state-of-art cylindrical RBC cell 30 cm both in height and diameter, designed and built such that its structure, placement in the cryostat as well as mechanical and electrical connections affect the confined RBC flow as little as possible. Still, it is evident that various corrections of the raw data are due, especially at high Ra $\gg 10^{10}$. 

Additionally, we have used one particular data series to discuss uncertainties connected with measurements of heat input, pressure and temperatures reads by sensors in various places inside and outside the RBC cell, as well as with the physical properties of working fluids -- cryogenic $^4$He gases -- relevant to the choice of the working points in the $p-T$ diagram, evaluated using available data bases \cite{REFPROP,HEPAK,XHEPAK}. In this study, we refrain from discussing the underlying physics of the high Ra RBC flow, which will be published elsewhere, but these investigations point to the difficulties in claiming any particular scaling at very high Ra. It is evident, however, that the discussed  uncertainties and corrections do not alter the overall character of the Nu/Ra$^{1/3}(\rm{Ra})$ versus Ra scaling. 

We believe that this study is useful and timely and will stimulate further improvements in investigations of the high Ra RBC flow, a model system for broad range of buoyancy-driven industrial and natural flows. 

\subsubsection*{Acknowledgements}
This research was supported by Czech Science Foundation (GACR) under  25-16812S. Fruitful discussions with  K. R. Sreenivasan and J. Krempl are warmly acknowledged.

\section*{Declarations}
The authors have no competing interests to declare.


\end{document}